\documentstyle[twocolumn,aps,amsmath,amssymb]{revtex}

\begin{document}

\title{Degenerate ground states and stability of spatial solitons in quasi-one-dimensional Bose-Einstein Condensates. }
\author{A. M. Dikand\'e} 
\address{Centre de Recherches sur les Propri\'et\'es \'Electroniques des Mat\'eriaux Avanc\'es, Facult\'e des Sciences,
  Universit\'e de Sherbrooke J1K-2R1 Sherbrooke-Qu\'ebec, CANADA \\ electronic mail: amdikand@physique.usherb.ca}

\date{\today}

\maketitle

\begin{abstract}
By examining the bound-state spectra of transverse fluctuations about one-dimensional, spatially localized dark and bright soliton 
wavetrains of the Gross-Pitaevskii equation, it is established that the low-temperature ground states of repulsive 
and attractive quasi-one-dimensional Bose-Einstein condensates are degenerate. In the one-soliton limit, both ground states 
are shown to possess two distinct transverse fluctuation modes which can couple to the spatial soliton: the first involves zero-energy exchange, costing 
only the soliton shape dressing via uniform translation of its centre of mass. The second mode contributes in a negative energy for the repulsive case,
but positive energy in the attractive case. This unstability of the repulsive spatial soliton against localized transverse fluctuation modes invalidates 
the Gross-Pitaevskii equation for stationary solitonic processes in repulsive 1D Bose-Einstein-Condensate systems. 
\end{abstract}


It was known for long time that dilute systems of Bose gases can become unstable against various kinds of interactions. However, 
the recent observations~\cite{1,2,3,4} of collective excitations in Bose-Einstein Condensed(BEC) gases motivated pressing renewal 
of interest to the problem in several, yet envisaged theoretical aspects. Solitonic behaviours~\cite{5} are thus currently being 
the most wanted perspectives in the light of recent experiments~\cite{1,2,3,4,6}. The framework is the Mean-Field(MF)~\cite{7,8}  
theory which provides acceptable mathematical tools to deal with the many-body character of the interacting system. The underlying 
concept sits in the possibility to cast a collection of microscopic processes to a macroscopic scale. At this scale, the order parameter relies 
on the macroscopic density of the excited atoms then suggesting their "quasi-particle"-like properties as they interact with the MF(internal) 
and trap(external) environments. Depending on the atomic species, the s-wave scattering length for atomic collisions will be positive~\cite{9} 
or negative~\cite{10}, such that the MF environment is either repulsive(self-defocusing effect) or attractive(self-focusing effect) vis-\`{a}-vis the Bose quasi-particle. 
The first effect promotes collective excitations in terms of "quasi-particle holes" with shape preserving macroscopic density. As for the second,
 it gives rise to quasi-particles in form of well localized hump whose macroscopic density also is shape preserving. These two different quasi-particle
 architectures are reproductible in the order parameter via the Gross-Pitaevskii(GP) equation~\cite{7,8} governing the dynamics of the ground-state wavefunction of a one-component BEC system. 
Concerning the true nature of this ground state, several questions are still seeking answers but however, there are serious indications of a dominant regime where the condensate tends to be frozen and thus oscillate rather slowly. This suggests the emergence 
of spatially localized collective excitations in the condensate~\cite{10}, the dynamics of which are well approximated in the MF description by a 
Hatree-Fock treatment~\cite{11} and which, in the contexts of one-dimensional(1D) and quasi-1D(Q1D) BEC systems, can readily be looked on as stationary states~\cite{12,13,14} of the GP equation. 
Regardless of dimensionality, the GP equation is comparable to a ND nonlinear Schr\"odinger(NDNS) equation with the trap-field 
potential playing role of perturbation. The NDNS equation(actually the 1D version) is popular in the literature of optical communication 
systems~\cite{15} for the richness of its soliton solutions: dark and bright solitons of this equation are representives of self-defocused
 and self-focused excitation modes, respectively.  \\ 
 To a viewpoint of the general approach to perturbed 1DNS equations~\cite{16}, their equivalence with the GP equation means that at least at low enough temperatures, the trap does not interfer
 in the initial stage of the soliton creation but instead, acts as a drive tightly confining the quasi-particle motion in a given direction along wich the 
soliton gets accelerated, increasing its lifetime. To this point, it may be instructive to draw attention on the fact that most of the currently reported typical quasi-particle 
lifetimes are short enough compared to the typical experimental time scales. Therefore to allow direct observations of their oscillations over 
controllable time periods, it is useful to increase their lifetimes by microtrap techniques~\cite{17}, 
hollow-blue-detuned laser beam traps~\cite{18}, and so on. On the other side, still in the spirit of the perturbation theory fo 1DNS equations, one should admit that 
effects of the trap can be advantageous in stabilizing the soliton wave-shape profile as shown in recent attempts assuming nonlinear periodic potential 
traps~\cite{14}. Moreover, they can cause soliton decays into distinct structures~\cite{4}. However these effects fall out of our context and will not be considered. Instead we  
assume an homogeneous potential field that allows us shifting the zero-point energy up to the desired characteristic energy scale.  \\
 The recent works on the solitonic properties of BEC systems focused mainly on the pure 1D regime(assuming that the characteristic length scale of the condensate in a 
given direction is far greater compared to the two others). A priori, a pure 1D description is unadequate and gives a rather oversimplified view of the physics. Instead the idea of 
a Q1D process fits better with experiments~\cite{19,20,21,22}. One of the relevant peculiarities of the 2D context is the possible coupling of transverse 
fluctuations to the confined 1D collective low-frequency oscillations in the ground-state phase of the condensate~\cite{22}. In view of the important role played by fluctuations 
in the occurence of a true condensate phase transition, it is surprising that only little attention has so far been devoded to the question. The few previous descriptions of transverse 
fluctuations in Q1D and Q2D BEC systems~\cite{22} postulated the interacting gas to be effectively kinematically $2D$, but undergoing zero-point oscillations in the transverse(frozen) direction 
while the collective motion of the condensate is along the confinement. The asserted goal of this assumption is quite appealing since it questions the possibility for the condensed ground state to be degenerate. 
However, a preriquisite to this question is the ones about the structural(wave shapes) and stability properties of the modes contributing to this degeneracy. 
Current stability theory for 1D soliton systems can provide relevant insights to these last questions. In this respect, we solved the Q1D GP equation including
 the coupling between the confined 1D collective condensate excitations and the transverse fluctuation modes taking burn in reponse to the creation of 
these 1D collective excitations. We found a rather rich bound-state spectrum for both repulsive and attractive systems which indicate ground states that are indeed degenerate. However, in the same 
time we observed that the current model could not be valid for at least one of the two regimes and hence need be reformulated. \\
In the first step we give a brief, but self-consistent summary of the general approach to the soliton stability in the presence of phonon fluctuations. 
This theory is specific to the current problem since we are interested in periodic spatial solitons~\cite{14,23}. In the next step, the theory will be 
applied to the repulsive and attractive BEC systems within the GP equation. Extension to the single-soliton limits will reveal itself more interesting since correspond to most of the current considerations. \\
  Starting, consider an eigenvalue problem describing the scattering of a quasi-particle of finite energy and mass, in a 1D nonlinear periodic potential:
\begin{equation}
\left[\frac{\partial^2}{\partial y^2} + P(\kappa,\epsilon) - n(n+1)\kappa^2sn^2(y)\right]f= 0 \label{a1}
\end{equation}
$P(\kappa,\epsilon)$ depends on the eigenvalue $\epsilon$ associate to the eigenfunction $f(y,\kappa)$, and $sn$ is the Jacobi Elliptic function of real 
modulus $\kappa$ between $0$ and $1$. The limits $\kappa=1$ and $n=1,2$ are well known as "Goldstone-Jackiw" eigenvalue problems~\cite{24}. For $\kappa\neq 1$ 
and integer $n$, equation (\ref{a1}) turns to a $n^{th}$-order Lam\'e equation~\cite{25} whose bound state spectrum has already been 
discussed~\cite{26,27} for $n=1$ and $n=2$. We improve these previous methods by introducing more simplified considerations leading straightly to the eigenvalues and 
eigenfunctions. In connection with the physical problem to be considered below, we will focus on the case $n=2$. In general, however, for even values of the integer $n$ the bound state spectrum 
will possess $2n+1$ eigenstates corresponding to the $2n+1$ possible values of $P(\kappa,\epsilon)$. Their wavefunctions are all real in $y$ and obey the relation:
\begin{equation}
f_(y,\kappa)= W(y,\kappa)sn^s(y)cn^c(y)dn^d(y) , \hspace{.1in}   (s, c, d)= 0, 1. \label{a2}
\end{equation}
$W(y,\kappa)$ is a polynomial in $sn^2(y)$ of degree $(n-s-c-d)/2$. Any two combinations of the nonzero values of $s$, $c$ and $d$ will give an eigenfunction of the eigenvalue 
equation provided the total degree of $f(y,\kappa)$ remains equal to $n$. The eigenfunctions for $s=c=d=0$ are obtained by writing down the polynomial $W(y,\kappa)$ as:
\begin{equation}
W(y,\kappa)= \sum^{\infty}_{r=0}\lambda_{2r}sn^{2r}(y) \label{a3}
\end{equation}
The sum in (\ref{a3}) runs over $N(=n/2)$ such that coefficients $\lambda_{2r}$ having indices in between $0$ and $N$ are all nonzero, and those of indices 
greather than $N$ are all zero. The compatibility relations obtained by replacing all the wavefunctions in equation (\ref{a1}) form a $(2n+1)^2$ matrix 
equation, the $2n+1$ eigenvalues of the matrix being the $2n+1$ allowed values of $P(\kappa,\epsilon)$. Thus constructed, the discrete spectrum of the eigenvalue problem 
(\ref{a1}) forms an orthonormal basis on the associate wavefunctions. \\
 Let us now see how this theory can help us solving the problem of bound fluctuation modes in Q1D BEC systems. Assuming spatially homogeneous 
potential trap $\mu_o$, the GP equation turns to the perturbed 2DNS equation i.e,
\begin{eqnarray}
i\hbar\frac{\partial}{\partial t}\psi(\vec{r},t)= - \frac{\hbar^2}{2m}\Delta\psi(\vec{r},t) + \mu_o\psi(\vec{r},t) + \nonumber \\ + g\mid \psi(\vec{r},t) \mid^2 \psi(\vec{r},t) \label{a4}
\end{eqnarray}
with $\psi(\vec{r},t)$ the macroscopic order parameter of the condensate and $\Delta$ the 2D Laplacian. The paramter $g$ is the strength of the MF interaction. 
It is customary to write $g= \frac{4\pi \ell}{m}$ where $\ell$ is the s-wave scattering length for inter atomic collisions and $m$ the mass of atoms. Positive and negative values of $\ell$ correspond 
respectively to a self-defocusing effect and a self-focusing effect of the medium on the BEC quasi-particle. We are interested in 1D collective excitations 
undergoing small-amplitude shape dressings induced by transverse fuctuations that couple to their spatial equilibrium positions~\cite{28,29} along the confinement. In the coherent regime, 
the macroscopic wavefunction describing such dressed solutions can be represented by:
\begin{equation}
\psi(\vec{r},t) =\sqrt{\rho_o} \left[\psi^{\mu}(x) + \varphi^{\mu}(x)\cos(\frac{p_\perp y}{\hbar})\right]\exp-(\frac{\imath\mu t}{\hbar} +\theta_o) \label{a5}
\end{equation}
$\rho_o$ is the uniform atomic density in the condensed phase, $\psi^{\mu}$ is the component of the macroscopic order parameter describing pure 1D coherent quasi-particle excitations with ernergy 
$\mu$ and we assume the condensate phase at equilibruim $\theta_o$. The function $\varphi^{\mu}$ represents the spatial dressing 
experienced by the wavefunction $\psi^{\mu}$ when transverse fluctuations $p_\perp$ of the amplitude of the order paramater are taken into consideration. The form (\ref{a5}) of the wavefunction solution
 is appropriate to describe 1D confinements in the Q1D system. Inserting (\ref{a5}) in equation (\ref{a4}) and keeping only linear terms in the fluctuations, 
we arrive at the following system of coupled equations:
\begin{equation}
\frac{\hbar^2}{2m} \frac{\partial^2}{\partial x^2}\psi^{\mu}(x)+\Delta\mu\psi^{\mu}(x) - g \rho_{o}\mid \psi^{\mu}(x) \mid^2 \psi^{\mu}(x)= 0 \label{a6}
\end{equation}
\begin{eqnarray}
\frac{\hbar^2}{2m} \frac{\partial^2}{\partial x^2}\varphi^{\mu}(x)+\Delta\mu\varphi^{\mu}(x) - 3g \rho_{o}\mid \psi^{\mu}(x) \mid^2 \varphi^{\mu}(x) \nonumber \\ = \frac{p^{2}_\perp}{2m} \varphi^{\mu}(x)  \label{a7}
\end{eqnarray}
Where we define the shifted zero-point energy as $\Delta\mu= \mu - \mu_o$. In the presence of the MF interaction, The order of magnitude of this zero-point energy is determined by the requirement that in the limit 
$t \rightarrow \infty$ the solution of the GP equation (\ref{a6}) effectively describes the true equilibrium state of an interacting Bose gas. That is, neglecting all fluctuation terms the wavefunction (\ref{a5}) must 
be representative of a complex order parameter with amplitude and phase which are macroscopically well defined at the equilibrium. Therefore it is only with time evolution and spatial modulations that the amplitude and phase 
of the order parameter will start fluctuating. In general at low dimensions these fluctuations lead to short-range correlations indicating the formations of topological defects. However, by the hierarchy of times scales and characteristic length scales involved 
with fluctuation phenomena in BEC systems, these topological defects are rather various and relate to both amplitude and phase fluctuations. Nevertheless, the relaxation of phase fluctuations generally requires longer time such that if one was 
interested in 1D short-range processes governed by an order parameter varying in time and space like (\ref{a5}), according to equation (\ref{a6}) the MF value $g\rho_o$ will constitue an appropriate energy scale. It turns out that the 
zero-point energy $\Delta\mu$ is about this scale. With these considerations, we can readily define the characteristic lengths $\zeta^{\pm}_x$ for the 1D coherent excitations of the order-paramater amplitude within 
the GP equation (\ref{a6}) either in terms of the zero-point energy, or in terms of MF energy scale. We consider the form:
\begin{equation}
\zeta^{(\pm)2}_x= \frac{\hbar^2}{2m\Delta\mu} \label{a8}
\end{equation}
With help of the incertainty principle, we can also introduce a characteristic time scale:
\begin{equation}
\tau_c= \frac{1}{\mid \Delta\mu \mid} \label{a9}
\end{equation}
As the energies of all the macroscopic processes governed by equation (\ref{a6}) are now contrained to the MF scale, it is natural to expect topological defects. With respect to the two signs of the MF strength $g$, two 
possible nonlinear topological solutions of this equation are:
\begin{equation}
\psi^{\mu +}(x)= \psi^{\mu +}_o(\kappa)sn\left(\frac{x}{\sqrt{1+\kappa^2}\zeta^+_x}\right)    
\end{equation}
\begin{equation}
\zeta^+_x= \frac{\hbar}{\sqrt{2m\Delta\mu}}, \hspace{.1in} \psi^{\mu +}_o(\kappa)= \sqrt{\frac{2g\rho_o}{\Delta\mu}}\frac{\kappa}{\sqrt{1+\kappa^2}}   \label{a10}    
\end{equation}
\begin{equation}
\psi^{\mu -}(x)= \psi^{\mu -}_o(\kappa)cn\left(\frac{x}{\sqrt{2\kappa^2-1}\zeta^-_x}\right)    
\end{equation}
\begin{equation}
\zeta^-_x= \frac{\hbar}{\sqrt{2m \mid \Delta\mu \mid}}, \hspace{.1in} \psi^{\mu -}_o(\kappa)= \sqrt{\frac{2g\rho_o}{\Delta\mu}}\frac{\kappa}{\sqrt{2\kappa^2-1}}  \label{a11}  
\end{equation}
Parameters $\kappa$ are such that $0< \kappa \leq 1$ for (\ref{a10}) and $\frac{1}{\sqrt{2}} < \kappa \leq 1$ for (\ref{a11}). $\Delta\mu$ and $g$ are both positive in (\ref{a10}) but negative in (\ref{a11}). For $0 < \kappa \leq 1$, (\ref{a10}) refers to a 1D periodic wavetrain of kinks. 
The kink sizes in the wavetrain are all equal to $\zeta^+_x \sqrt{1+\kappa^2}$, and the wavetrain period is: 
\begin{equation}
L_+= \sqrt{1+\kappa^2}K(k)\zeta^+_x \label{a12}
\end{equation}
where K(k)(and E(k) hereafter) is the Complete Jacobi Elliptic Integral. In the limit $\kappa \rightarrow 0$, the kink wavetrain decays into harmonic oscillations whereas 
in the opposite limit i.e.$\kappa \rightarrow 1$, it collapses turning to a single kink. The single kink describes a localized spatial density hole in the ground state of the repulsive BEC system. 
Remark that in this last limit the period $L_+$ becomes infinite. It is interesting to 
calculate the number of particles in this kink wavetrain. We note from (\ref{a10}) that the wavetrain function vanishes at $x=0$ while being maximum, 
i.e. $\psi^{\mu +}(x)= \psi^{\mu +}_o(\kappa)$ at each period $L_+$. In virtue of this periodicity we can readily shift the position of the wavetrain so that 
its density becomes maximum at its centre of mass. By this shift not only we avoid processes involving unsaturable particle densities, but also we provide the wavetrain with a finite-energy rest frame. 
This leads to the following expression of the particle number:
\begin{eqnarray}
N^+(\kappa)= \frac{\rho_o \sqrt{1+\kappa^2}\psi^{\mu(+)2}_o(\kappa)\zeta^+_x}{\pi\kappa^2}\left[E(k) - \kappa^2_1 K(k)\right], \nonumber \\    \kappa^2_1= 1- \kappa^2 \label{a13}
\end{eqnarray}
This particle number in the kink wavetrain carries a total energy amounting to:
\begin{eqnarray}
E^+(\kappa)= \sqrt{\frac{2\hbar^2}{9\pi^2 g^2 m(1+\kappa^2)}}[E(k)- \nonumber \\ - {\frac{\kappa^2_1}{1+\kappa^2}} K(k)](\mu - \mu_o)^{3/2} \label{a14}
\end{eqnarray}
This is precisely the amount of energy needed to create the periodic wavetrain of kink condensate. \\
Similarly, for $\frac{1}{\sqrt{2}} < \kappa \leq 1$, (\ref{a11}) will describe a 1D periodic wavetrain of pulse(envelope) structures, the mean sizes of which are $\zeta^-_x \sqrt{2\kappa^2-1}$. 
The period of this envelope wavetrain is:
\begin{equation}
L_-= \sqrt{2\kappa^2-1}K(k)\zeta^+_x \label{a15}
\end{equation}
and its particle number and energy are respectively:
\begin{equation}
N^-(\kappa)= \frac{\rho_o \sqrt{2\kappa^2-1}\psi^{\mu(-)2}_o(\kappa)\zeta^-_x}{\pi\kappa^2}\left[E(k) - \kappa^2_1 K(k)\right] \label{a16}
\end{equation}
\begin{eqnarray}
E^-(\kappa)= \sqrt{\frac{2\hbar^2}{9\pi^2 g^2 m(2\kappa^2-1)}}[{\frac{\kappa^2_1}{2\kappa^2 -1}}K(k)+ \nonumber \\ + 6\kappa^2 E(k)](\mu_o - \mu)^{3/2} \label{a17}
\end{eqnarray}
As the first, this solution also reduces to harmonic oscillations in the limit $\kappa\rightarrow 0$ but transforms to a single envelope soliton as $\kappa
\rightarrow 1$. This single envelope soliton describes a localized spatial density hump forming in the ground state of the system in the attractive regime. \\
We now turn to equation (\ref{a7}) describing the spatial dressing. We set $\varphi^{\mu}=\varphi^{\mu}_o f^{\mu}(x)$, where 
$f^{\mu}(x)= f^{\mu\pm}(x)$ and $\varphi^{\mu}_o(x)= \varphi^{\mu\pm}_o= \sqrt{\frac{g\rho_o}{\Delta\mu}}$. In terms of the dimensionless fonctions $f^{\mu}$, equation (\ref{a7}) becomes:
\begin{equation}
\left[\frac{\partial^2}{\partial z^2} + P(\kappa,p_{\perp}) - 6\kappa^2sn^{2}(z) \right]f^{\mu}(z)= 0 \label{a18}
\end{equation}
This is nothing but the eigenvalue equation (\ref{a1}) with $n=2$. As functions of the physical parameters in the present model, we have:  
\begin{equation}
z_+= \frac{x}{\sqrt{1+\kappa^2}\zeta^+_x}, \hspace{.1in}   z_-= \frac{x}{\sqrt{2\kappa^2-1}\zeta^-_x}  \label{a19}
\end{equation}
\begin{eqnarray}
P_+(\kappa,p_{\perp})= (1-\frac{p^2_{\perp}\zeta^{(+)2}_x}{\hbar^2})(1+\kappa^2), \nonumber \\   P_-(\kappa,p_{\perp})= 4\kappa^2+1-\frac{p^2_{\perp}\zeta^{(-)2}_x(2\kappa^2-1)}{\hbar^2} \label{a20}
\end{eqnarray}
In terms of these parameters, the bound state solutions of (\ref{a18}) follow from the method presented above. In both cases they read by turns:
\begin{equation}
f^{\mu}_o(x)= A_o(\kappa)cn(z)dn(z), \hspace{.1in}   P_o(\kappa, p_{\perp})= 1+\kappa^2  \label{a21}
\end{equation}
\begin{equation}
f^{\mu}_1(x)= A_1(\kappa)cn(z)sn(z), \hspace{.1in}   P_1(\kappa, p_{\perp})= 4+\kappa^2  \label{a22}
\end{equation}
\begin{equation}
f^{\mu}_2(x)= A_2(\kappa)sn(z)dn(z), \hspace{.1in}   P_2(\kappa, p_{\perp})= 1+4\kappa^2  \label{a23}
\end{equation}
\begin{equation}
f^{\mu}_{(3,4)}(x)= A_{(3,4)}(\kappa)\left(sn^2(z) - \frac{1+\kappa^2}{3\kappa^2} \pm \frac{\sqrt{1-\kappa^2 \kappa^2_1}}{3\kappa^2}\right)
\end{equation}
\begin{equation}
P_{(3,4)}(\kappa, p_\perp)= 2(1+\kappa^2) \mp \frac{\sqrt{1-\kappa^2\kappa^2_1}}{2}  \label{a24}
\end{equation}

Tables \ref{tab:table1} and \ref{tab:table2} list exact expressions of $A^{(\pm)}_i$ and $p^{(\pm)2}_{\perp, i}$.

Parameters in these tables are calculated using the ortho normalization relation for the Jacobi Elliptic functions \cite{27}. We find:
\begin{equation}
a_o= (1+\kappa^2)E(\kappa) - \kappa^2_1 K(\kappa) \label{a25a}
\end{equation}

\begin{equation}
a_1= (1+\kappa^2_1)E(\kappa) - 2\kappa^2_1 K(\kappa) \label{a25}
\end{equation}
\begin{equation}
a_2= (2\kappa^2 -1)E(\kappa) + \kappa^2_1 K(\kappa) \label{a26}
\end{equation}
\begin{eqnarray}
a_{(3,4)}= \left({\frac{2}{3\kappa^2}}\right)^2 \delta \left[\delta K(\kappa) \mp 3(K(\kappa) - E(\kappa)) \pm \right. \nonumber \\ \left. \pm (1+\kappa^2)K(\kappa)\right], \hspace{.1in} \delta^2= 1-\kappa^2\kappa^2_1  \label{a27}
\end{eqnarray}

We end with a discussion on the asymptotic behaviours of the essential parameters calculated above when $\kappa= 1$. For the repulsive regime, 
the kink wavetrain solution (\ref{a10}) will become:
\begin{equation}
\psi^{\mu +}(x)= \psi^{\mu +}_o(1)\tanh\left({\frac{x}{\sqrt{2}\zeta^{+}_x}}\right), \hspace{.1in}  \psi^{\mu +}_o(1)= \sqrt{{\frac{g\rho_o}{\Delta\mu}}}  \label{a28}    
\end{equation}
and the associate bound states reduce to:
\begin{equation}
f^{\mu}_o(x)= A_o(1)sech^2({\frac{x}{\sqrt{2}\zeta^{+}_x}}),  \hspace{.1in}  p^2_{\perp,o}= 0 \label{a29}    
\end{equation}
\begin{eqnarray}
f^{\mu}_1(x)=  A_1(1)sech({\frac{x}{\sqrt{2}\zeta^{+}_x}})\tanh({\frac{x}{\sqrt{2}\zeta^{+}_x}}), \nonumber \\ f^{\mu}_1(x)= f^{\mu}_2(x),  \hspace{.1in} p^2_{\perp,1}= p^2_{\perp,2}= {\frac{-3\hbar^2}{2\zeta^{(+)2}_x}}   \label{a30}    
\end{eqnarray}
\begin{equation}
f^{\mu}_{3,4}(x)= 0  \label{a31}
\end{equation}
As for the attractive case, the envelope-soliton wavetrain obtained in (\ref{a11}) becomes:
\begin{equation}
\psi^{\mu -}(x)= \psi^{\mu -}_o(1)sech\left({\frac{x}{\zeta^{-}_x}}\right), \hspace{.1in}  \psi^{\mu -}_o(1)= \sqrt{{\frac{g\rho_o}{\Delta\mu}}}  \label{a32}    
\end{equation}
while the associate bound states reduce to:
\begin{equation}
f^{\mu}_o(x)= A_o(1)sech^2({\frac{x}{\zeta^{-}_x}}),  \hspace{.1in}  p^2_{\perp,o}= {\frac{3\hbar^2}{2\zeta^{(-)2}_x}}   \label{a33}    
\end{equation}
\begin{eqnarray}
f^{\mu}_1(x)=  A_1(1)sech({\frac{x}{\zeta^{-}_x}})\tanh({\frac{x}{\zeta^{-}_x}}), \nonumber \\ f^{\mu}_1(x)= f^{\mu}_2(x),  \hspace{.1in} p^2_{\perp,1}= p^2_{\perp,2}= 0  \label{a34}  
\end{eqnarray}
\begin{equation}
f^{\mu}_{3,4}(x)= 0  \label{a35}
\end{equation}
The meaning of these results is that for each regime there is at least one transverse fluctuation mode that couples to the single-soliton without energy exchange. 
This zero-energy mode then contributes in increasing the soliton stability, costing only a shape correction via uniform translation of the soliton centre of mass. 
The second bound state of the repulsive regime has negative energy and implies cost of energy to the soliton, which sees its creation energy lowered thus becoming unstable. 
On the contrary, the non-zero bound-state mode of the attractive regime is positive and then contributes to the stability of the bunch. In fact these two distinct behaviours can indicate which of the 
two regimes is properly described within the 1DNS equation. From what precedes, the repulsive GP equation gives and unstable 1D spatial soliton phase and therefore is unadequate for the study of the spatial 
localized density-hole excitations(dark solitons) in the ground state of the Q1D repulsive BEC system. It is instructive remarking that several other authors recently raised the
same question \cite{30,31}.  
The values of the particle numbers in this limit also can be helpful for checking the consistency of the theory. In the repulsive 
case the quantity (\ref{a13}) gives $N^+(1)= \rho_o d_K$, where $d_K= \sqrt{2}\zeta^+_x$ is the actual width of the single kink resulting from (\ref{a10}). As for the attractive 
case expression (\ref{a17}) reduces to $N^-(1)= 2\rho_o \zeta^-_x$. The fact that the number of particles in the one-soliton condensate limit is the product of the condensate atomic density time the soliton 
size, agrees with the characteristic length scales defined above: rescaling the x axis in units of the coherence lengths $\zeta^{\pm}_x$ gives the soliton densities(but no more the particle number in the soliton) $\rho_o$ and 
$2\rho_o$ for the two respective regimes. 

\begin{acknowledgments}

Part of this work was done during visits at the Abdus Salam International Centre for Theoretical Physics(Trieste-Italy). 
The author is endebted to S. R. Shenoy for continuous support and advices. 

\end{acknowledgments}

\newpage

\begin{table} 
\caption{\label{tab:table1} Amplitudes $A^{(\pm)}_{i=0,1,2,3,4}$ of the five bound states: $d^+_{\kappa}= \sqrt{1+\kappa^2}$ and $d^-_{\kappa}= \sqrt{2\kappa^2-1}$. The $a_{i}$ are defined in the text. } 
\begin{tabular}{ccccc}
 regime&$A^2_o$ &$A^2_1$ &$A^2_2$& $A^2_{(3,4)}$\\ \hline \\
repulsive& $\frac{3\pi\kappa^2}{4a_o\zeta^+_{x} d^+_{\kappa}}$ & $\frac{3\pi\kappa^4}{4a_1\zeta^+_{x} d^+_{\kappa}}$  & $\frac{3\pi\kappa^2}{4a_2\zeta^+_{x} d^+_{\kappa}}$ & $\frac{\pi}{2a_{(3,4)}\zeta^+_{x} d^+_{\kappa}}$ \\ \\ 
attractive& $\frac{3\pi\kappa^2}{4a_o\zeta^-_{x} d^-_{\kappa}}$ & $\frac{3\pi\kappa^4}{4a_1\zeta^-_{x} d^-_{\kappa}}$  & $\frac{3\pi\kappa^2}{4a_2\zeta^-_{x} d^-_{\kappa}}$ & $\frac{\pi}{2a_{(3,4)}\zeta^-_{x} d^-_{\kappa}}$ \\ \\
\end{tabular}
\end{table}

\begin{table} 
\caption{\label{tab:table2} Eigenvalues $p^{(\pm)2}_{\perp,i=0,1,2,3,4}$ of the five bound states. } 
\begin{tabular}{ccccc}
 regime& $p^2_{\perp, o}$& $p^2_{\perp, 1}$& $p^2_{\perp, 2}$ & $p^2_{\perp, (3,4)}$ \\ \hline  \\
repulsive& $0$ &$\frac{-3\hbar^2}{\zeta^{(+)2}_{x}d^{(+)2}_{\kappa}}$ &$\frac{-3\hbar^2\kappa^2}{\zeta^{(+)2}_{x}d^{(+)2}_{\kappa}}$  & $\frac{-\hbar^2}{\zeta^{(+)2}_{x}}\left[1\pm \frac{\sqrt{\delta}}{2d^{(+)2}(\kappa)}\right]$ \\ \\
attractive& $\frac{3\hbar^2\kappa^2}{\zeta^{(-)2}_{x}d^{(-)2}_{\kappa}}$ &$\frac{-3\hbar^2\kappa^{2}_1}{\zeta^{(-)2}_{x}d^{(-)2}_{\kappa}}$ & 0& $\frac{\hbar^2}{\zeta^{(-)2}_{x}}\left[1\mp \frac{\sqrt{\delta}}{2d^{(-)2}(\kappa)}\right]$ \\ \\
\end{tabular}
\end{table}

\end{document}